\documentclass[11pt,a4paper]{article}
\usepackage[left=15mm, top=10mm, right=15mm, bottom=20mm, nohead=true, nofoot=true]{geometry}

\usepackage[utf8]{inputenc}
\usepackage{authblk}
\usepackage{indentfirst}
\usepackage{misccorr}
\usepackage{graphicx}
\usepackage{amsmath}
\usepackage{amssymb}
\usepackage{multicol}
\usepackage{bm}
\usepackage{longtable}
\usepackage{pdflscape}
\usepackage{epstopdf}
\usepackage{multirow}
\usepackage{adjustbox}
\usepackage{amsfonts}
\usepackage{ifthen}
\usepackage{fancyhdr}
\usepackage{setspace}
\usepackage{xcolor}
\usepackage[round,authoryear,comma,sort&compress,numbers]{natbib}
\usepackage[toc,title,page]{appendix}
\usepackage[hidelinks]{hyperref}
\usepackage{url}

\hypersetup{
    colorlinks=true,
    linkcolor=green,
    citecolor=blue,
    filecolor=magenta,
    urlcolor=cyan,
    pdftitle={Sharelatex Example},
    pdfpagemode=FullScreen,
}

\fancypagestyle{plain}{\chead{\texttt{\textcolor{brown}{}}}}
\title{\textbf{Dynamics of clusters of  galaxies in the presence of dark energy: Virgo cluster}}
\author[]{H.A. Harutyunian  \thanks{hhayk@bao.sci.am}}
\author[]{E.H. Nikogosyan }
\affil[]{\scriptsize Byurakan Astrophysical Observatory, Armenia}

\begin{document}
\pagestyle{empty}
\newpage
\pagestyle{fancy}
\label{firstpage}
\date{}
\maketitle

\begin{abstract}
   It is shown that, owing to the interaction of baryonic matter with the carrier of dark energy, all configurations of baryonic matter acquire energy and inevitably must expand. This conclusion applies to all hierarchical levels of the baryonic universe, including galaxy clusters. We propose a simple statistical method for identifying possible radial motions of galaxies within clusters. To illustrate this method, we examined the structural features of the Virgo galaxy cluster and identified its substructure, comprising groups of galaxies of varying multiplicity. Galaxies in the substructure are somewhat brighter than those in the overall cluster, and each subgroup contains an active galaxy. Subgroups are considered to be the product of primordial ejections of matter from the central generator galaxy. It is shown that the average stellar magnitude of galaxies in subgroups positively correlates with their average velocity. This correlation can be interpreted as evidence of cluster expansion.
\end{abstract}

\section{Introduction}
The clusters of galaxies are one of the highest hierarchical levels of baryonic configurations yet accessible for easy identification. Searching for general regularities in the formation and evolution issues at the mentioned level has always been one of the central problems of extragalactic cosmogony. Especially, the evolution path has been and remains one of the essential issues.

Modern views underlying clusters and galaxy formation theories have adopted a dualistic approach: it is believed that our baryonic Universe originated in a huge explosion and continues to expand, but all cosmic objects beginning with atomic nuclei, planetary systems, stars, and extending to galaxies and their clusters, are generally regarded as being in dynamical equilibrium, having reached this state through gravitational collapse. Most astrophysicists adhere to this classical concept of the formation of cosmic structures. 

However, this approach leads to some inconsistency when comparing the velocity fields and observed masses. This paradox was first observed for our galaxy and the Coma cluster of galaxies. The observed masses cannot provide a dynamical equilibrium for the measured velocity dispersion. Therefore, in the 30s of the last century, Fritz Zwicky proposed the well-known hypothesis of the existence of non-baryonic dark matter \citep{Zwicky1933, Zwicky1937ApJ}. It happened almost a century ago, and since then, dark matter has been used as a convenient free parameter to explain several unresolved problems. Later on, it became an unchangeable tool to interpret the’ rotation curves of galaxies as well. 

In the mid-20th century, Ambartsumian proposed an alternative concept, suggesting that decay and fragmentation processes dominate at all hierarchical levels of the baryonic Universe \citep{Ambartsumian1958, Ambartsumian1961}. Observational manifestations of his ideas, such as the presence of young stars and the activity of galactic nuclei, were incorporated into astrophysical research; however, his central idea related to decay processes was finally rejected.

At the end of the last century, dark energy was discovered. One might expect that this discovery could significantly change the situation. However, the prevailing paradigm remained strong. In this article, we assume that the unknown carrier of dark energy continuously interacts with baryonic matter and transfers a non-zero amount of energy to it. Moreover, this process occurs continuously at all hierarchical levels. Based on this, and applying the known laws of physics, we inevitably arrive at conclusions that confirm Ambartsumian's ideas, including the expansion of galaxy clusters \citep{Harutyunian2008, Harutyunian2024}.

Determining the expansion of galaxy clusters, therefore, remains a highly challenging issue. While radial velocities can be measured with reasonable precision, estimates of galaxy distances and their three-dimensional distribution within a parent cluster remain uncertain. Proper motions and tangential velocities are also poorly constrained.

\section{New player in the old game - dark energy.}

Before the end of the last century, researchers had been desperately looking for the deceleration rate of the Universe's expansion. Instead, the expansion appeared to be accelerating \citep{Riess1998, Perlmutter1999}. This unexpected discovery opened new perspectives for further analyzing astrophysical phenomena. 

Over the past quarter-century, dark energy, introduced into the astrophysical toolbox as an accelerating energy resource, has been intensively studied using both observational and theoretical methods. It is currently believed that dark energy accounts for approximately 70 percent of the total mass/energy of the Universe, while baryonic matter accounts for only 5 percent. This ratio suggests that dark energy, although we currently know nothing about its carrier, must somehow affect the baryonic matter and the evolution of baryonic configurations. Therefore, it is surprising that this issue has received almost no in-depth consideration.

Let us recall the discovery of this new type of energy: it was introduced into modern physics based on the observational detection of the acceleration of the Universe's expansion. So, an essential and intrinsic property of its carrier is its continuous interaction with baryonic matter, which is how galaxies acquire kinetic energy. Therefore, to construct a complete picture of the evolution of the baryonic world, all the consequences of this interaction and energy exchange must be analyzed in detail.

The modified Kant-Laplace mechanism remains the only mechanism to interpret the formation of all baryon configurations to this day. Therefore, all these structures are considered to be in equilibrium. Let us assume for a moment that the mentioned mechanism is liable.  Then, all baryonic structures are in dynamic equilibrium according to the dominant viewpoint, such that the virial theorem holds for all of them:

\begin{equation}
 W=2T+U,   
\end{equation}
where  $T$ and $U$ are the kinetic and potential energies, respectively. Evidently, for such systems, the total energy is negative:

\begin{equation}
E=T+U<0.
\end{equation}
By contrast, dark energy is positive by definition, since it accelerates galaxies and performs work. We thus have two interacting systems in any given volume: baryonic structures with negative total energy and the carrier of dark energy, which possesses positive energy. According to the second law of thermodynamics, energy flows from a system possessing positive energy to another system with negative energy. Therefore, dark energy is continuously transferred into baryonic structures, increasing their entropy.

Because dark energy is accepted to be uniformly distributed on all scales, this interaction should occur wherever baryonic matter exists. The amount of energy transferred to baryonic configurations per unit time is less important than the fact of the transfer itself, since energy is a cumulative substance and accumulates over time. Eventually, the virial equilibrium is disrupted due to this continuous transfer, and after a finite time, we have

\begin{equation}
W>0, 
\end{equation}
implying that the system expands at a nonzero velocity. One important note: as expansion continues, the interaction volume grows, thereby increasing the rate of energy input and further accelerating expansion.

This conclusion follows directly from accepted empirical facts and physical laws: the existence of dark energy, its interaction with baryonic matter, and the laws of thermodynamics. The accelerated expansion of the Universe on cosmological scales is the trivial manifestation of this process. On much smaller scales, three expanding systems have been identified within the Solar System: the pairs Earth-Moon \citep{Dickey1994}, Saturn-Titan \citep{Lainey2020a,Lainey2020b}, and the Sun-Earth \citep{Krasinsky2004}. The observed relative expansion rates ($\delta =\Delta R/R$) for these three systems and the Hubble expansion are $9.94\times {{10}^{-11}}$,  $9.42\times {{10}^{-11}}$,  $1.\times {{10}^{-12}}$, and $6.71\times {{10}^{-11}}$, respectively.

The similarity of the values for the two planet-satellite pairs, differing by only 5 percent, is absolutely striking, as is their approximate agreement with the Hubble expansion value (differing by 30 percent). Within the Solar System, tidal effects contribute to satellite recession, explaining the slightly higher expansion rates for orbital radii of moons compared to the Hubble value. The anomalous rate of increase in the Astronomical Unit, which is too large for traditional mechanisms alone and too small for purely cosmological ones, can also be reconciled if the same physical principles are applied self-consistently down to baryonic and nuclear scales \citep{Harutyunian2019}.

\section{Statement of the problem and observational data}

Measurements of radial velocities for large galaxy samples enable detailed studies of cluster dynamics and kinematics. A natural question arises: can the general kinematical properties of a cluster be determined solely from galaxy radial velocities? Specifically, can we determine whether a cluster is stationary and balanced or violates the virial theorem?

Active phenomena in galactic nuclei, such as nuclear ejections, explosions, and decay of nuclei, are frequently observed.  Ambartsumian was the first to emphasize the significance of these processes in the evolution of galaxies and to study the spatial distribution of galaxies in multiple systems to infer the direction of the evolutionary path \citep{Ambartsumian1958, Ambartsumian1961}. This was long before supermassive black holes became the standard explanation for the activity of galactic nuclei. A key feature of Ambartsumian’s concept is that galactic nucleus activity is intrinsic. His model was later rejected primarily because it implied the existence of stable clumps of superdense matter, deemed unphysical.

One of the present authors, in studying correlations between the properties of cD galaxies and their host clusters, concluded that such galaxies may act as generators of cluster formation \citep{Harutyunian2008}. This is consistent with the dark energy-baryon energy exchange model \citep{Harutyunian2019, Harutyunian2021, Harutyunian2024}, which predicts an eventual energy imbalance that requires the release of excess energy as radiation, baryonic structural fragments, and kinetic energy.

This approach implies that, in at least some clusters, a fraction of galaxies should be moving radially outward from the center where they were born. While this would be directly testable with accurate three-dimensional galaxy positions, such data remain elusive. However, a statistical method may be applied: both geometric effects and intergalactic absorption reduce the apparent brightness of galaxies lying farther along the line of sight. The question is whether the magnitude of this dimming is detectable and usable for kinematic studies.

These two effects differ fundamentally. Intergalactic absorption is independent of the cluster’s distance and depends only on its internal conditions. Observations indicate that intergalactic gas consists of a low-density, high-temperature plasma, with one atom per cubic meter, and relativistic electrons as the main scattering agents via the inverse Compton effect. This process attenuates light without distorting images.

The geometric effect, by contrast, depends on cluster distance and decreases as the distance increases. For two galaxies along the same line of sight at distances ${{r}_{1}}$ and ${{r}_{2}}$ and (${{r}_{1}}<{{r}_{2}}$), the distance modulus difference is: 
\begin{equation}
\Delta m=5\lg ({{{r}_{1}}}/{{{r}_{2}})=5\lg (1+{\Delta r}/{{{r}_{1}})}\;}, 
\end{equation}
where  $\Delta r={{r}_{2}}-{{r}_{1}}$. For distant clusters, the ratio ${\Delta r}/{{{r}_{1}}}\;$ tends to zero because $\Delta r$ is limited by the diameter of the cluster. Thus: 
\begin{equation}
\Delta m\approx 5{\Delta r}/{{{r}_{1}}}\, 		
\end{equation}
making the rate of the geometric effect's disappearance straightforward to estimate.

We therefore expect that, within the cluster, galaxies will become systematically fainter with increasing depth from the “leading edge”. The task is to search for a correlation between apparent magnitude and radial velocity. A positive correlation would be a necessary condition for ongoing cluster expansion. 

The observational material consists of published catalogs of the Virgo Cluster. Given the variety of objects within the cluster, the data are necessarily heterogeneous, having been obtained with different instruments and telescopes across multiple decades. As a result, the data vary in quality and precision. Nevertheless, thanks to the systematic efforts of numerous astronomers, the Virgo Cluster remains one of the best-studied clusters.

The Virgo Cluster is a prototype of an irregular cluster with a complex, multi-component structure. Within the framework of our cosmogonic approach, this means that the formation of galaxies in the cluster occurred through successive decays and outflows. Then, in addition to the primary generator of galaxies, which is M87 by all indications, there are undoubtedly other fairly powerful centers of galaxy formation. This complicates the overall picture, but if the basic concept of general expansion is correct, then at least at the statistical level, the "fingerprints" of the cluster-building mechanism should be revealed. 

We use data existing in the literature and open electronic catalogs. The basis for the observational material is the list published in 1985 \citep{Binggeli1985}, which contains data on 2096 galaxies from a region of the sky with an area of 140 square degrees centered at the coordinates $\alpha ={{12}^{h}}{{25}^{m}}$ and $\delta ={{13}^{o}}$. This list is complete up to $B={{18}^{m}}$. Radial velocities are given for 564 galaxies. In a later paper \citep{Binggeli1993}, the radial velocities of 144 additional galaxies were published. Besides, the data for 131 galaxies from the previous list were refined. The final list includes 708 galaxies with measured radial velocities, 522 of which were identified as cluster members. We are considering this list. Note that 30 out of 522 are radio and X-ray sources according to NED data.

We have constructed an image of the projected local density of the cluster (the reference point is the coordinates of M87) using the Dressler method |\citep{Dressler1980}. Its structure coincides with the general picture of X-ray radiation \citep{Bohringer1994}.

\section{Hierarchical structure of the cluster}

Considering the Virgo Cluster, we started by studying its structural features. We applied the Htree analysis method \citep{Serna1996} to study the hierarchical structure of the cluster.  For this purpose, the total energy estimates are implied to sort the objects by their degree of connectivity. 

This analysis revealed seven groups, four triplets, and five pairs of galaxies within the cluster. For the further examination, some physical parameters are determined, such as the number of galaxies in the subgroup; the average radial velocity in km/sec; its dispersion; the average intergalactic distance; the relative asymmetry, characterizing the degree of deviation of the radial velocity distribution from Gaussian (skewness – asymmetry, with ideal coincidence S=0); the excess, characterizing the ratio of the values of the radial velocity distribution function of galaxies in the “arms” of distribution to its central value (kurtosis – excess, in the ideal case K=0) \citep{Solanes1999}; and the probability of deviation of the projection positions of galaxies in the cluster from uniform \citep{Salvador-Sole1993}.

The data for the field, galaxies that are not part of the substructure (pairs from the field are not excluded), are presented. The bulk of the galaxies is included in the field; however, after excluding the substructure, the velocity dispersion decreases slightly. Moreover, the field has a more uniform distribution of object projections than the cluster as a whole.

The central pair includes the galaxy M87. This pair is located in the region of the density and the radial velocity distributions' maxima. The radial velocity distribution (RVD) was constructed using the ''wavelet'' method \citep{Fadda1998}. M87 itself is somewhat shifted from the main maximum of the cluster density, which is located in the direction of the galaxy’s jet \citep{Vaucouleurs1979}.

All subgroups coincide with the maxima of both the density and radial velocity distributions. Their location is in good agreement with the cluster’s X-ray density maxima \citep{Bohringer1994}. Binggeli et al \citep{Binggeli1987} suggested that the Virgo cluster consists of two main parts or clouds A and B, formed around the giant galaxies M87 and M49, respectively. Interestingly, the radial velocity of M87 also does not coincide with the average radial velocity of cloud A. However, cluster analysis did not reveal a structural division; the entire cluster is a single system, which includes subgroups. M49 itself is part of group 1 and is its dominant member. 

Our analysis once again confirms the viewpoint that the "clouds" of galaxies around M87 and M49 are not separate, interconnected formations, but represent a single system (see, for example, \citep{Nolthenius1993}. Another confirmation of such a structural feature is the distribution of radial distances of the cluster galaxies from M87. It coincides almost exactly with the general distribution of intergalactic distances. The probabilistic difference from the latter is only 6 percent. In other words, the projected arrangement of objects in the cluster is uniform with a probability of 94 percent (calculations were made with a resolution of 0.3 Mpc) \citep{Salvador-Sole1993}. Note that the radial velocity of M87 coincides with the average velocity of the entire cluster.

The objects of the substructure are, on average, brighter than those in the cluster as a whole. This may be a consequence of the selection method, since the galaxies' luminosity is one of the primary parameters in the analysis. However, it can also be a genuine physical feature of the substructures. It should be noted that, in terms of morphological composition, the substructure does not differ significantly from the overall cluster.

The 4 pairs include only 2 S0 galaxies, while the remaining 6 are spirals. Galaxies in each pair belong to the same morphological type. The morphological composition of the substructure is not uniform like that of the pairs. However, the lower the system multiplicity, the more uniform the substructures' morphological content. On the other hand, if the number of galaxies increases in the substructure patterns, their morphological distribution approaches the composition of the cluster. 

The distribution of radial velocity dispersion (DRVD) relative to the galaxies M87 and M49 is constructed using 500 arc-second intervals. DRVD is compiled relative to M87 for both the entire cluster and the field. The central galaxy is located at the minimum of gravitational potential, and the DRVD of this region has a classic peaked profile, typical for clusters with a central cD galaxy \citep{Hartog1996}. At a distance of 0.47 Mpc, it reaches a minimum. The DRVD of the entire cluster and the field do not differ practically. The presence of substructure in this case does not affect the distribution of the radial velocity dispersion.

\section{Relationship between average stellar magnitudes and average radial velocity}

We already mentioned in the previous paragraphs that some features of the DRVD can be interpreted within the framework of the cluster-building concept, suggesting successive fragmentation or a chain of decays and ejections of pre-galactic matter from the core of M87. 

On the other hand, the presence of successive fragmentation distorts the velocity field, transforming it from purely radial to a superposition of radial ones. Therefore, it seems very important that dwarf galaxies, distributed in the same region as elliptical and lenticular galaxies, exhibit much larger radial velocity dispersion. This can be easily explained if a part of these dwarf galaxies were ejected not directly from the primary generator, but from the nuclei of next-generation galaxies. Then, obviously, the velocity dispersion of this sample should be defined as the total dispersion of objects of the first and next generations. 

To avoid confusion introduced by galaxies of the next generation, we will consider only subgroups. Most likely, the subgroups have been formed from clumps of matter, ejected directly from the central galaxy. Most probably, the next generation objects could not possess such matter resources. From analysis of relevant data, one finds an obvious correlation between the average stellar magnitudes of the subgroups and their radial velocities: the fainter the object, the greater its average radial velocity. One can express this correlation in the form
\begin{equation}
{{V}_{r}}=537(\pm 308)m-5228(\pm 3804) 
\end{equation}
and find the correlation coefficient to be $R=0.48$. This is a very weak positive correlation. However, if one excludes the subgroup furthest from the center in the sky along the x-coordinate, then the relation (6) transforms into
\begin{equation}
{{V}_{r}}=1019(\pm 303)m-11057(\pm 3748)
\end{equation}
with a high correlation coefficient $R=0.81$. 

The corresponding diagram shows that the maximum difference between the stellar magnitudes is approximately 2m.5. A simple calculation using formula (4) allows us to estimate that the weakening due to the geometric effect cannot be greater than $0^{m}.5 - 0^{m}.8.$. So, we can conclude that the weakening by approximately $1^m.7 - 2^m.0$ occurs due to intergalactic absorption.

\section{Concluding remarks}

A self-consistent application of physical laws allows us to conclude that all baryonic structures continuously interact with the carrier of dark energy. Taking into account that all baryonic structures have negative total energy, while dark energy, conversely, is positive, we conclude that baryonic structures continually receive additional energy. Since energy is cumulative, the amount of energy transferred per unit time is irrelevant. What is important is that this energy is non-zero and accumulates over time in any baryonic structure.

The most obvious result of this interaction is the accelerating expansion of the universe, which allowed the discovery of dark energy. In the solar system, the expansion rate of Earth's orbit, as well as the orbits of the Moon and Titan, has been measured. Moreover, the latter two velocities, taking into account the scale factor, are almost identical to each other, are of the same order of magnitude as the analogous value calculated using the Hubble constant, and are only 25-30 percent greater.

Since the conclusion about the inevitability of expansion applies to all baryonic structures, we expect this effect to be observed in galaxy clusters as well. Therefore, we are investigating the nearest cluster in Virgo. The goal of the study is to determine whether there is a statistically significant correlation between galaxies' redshifts and their apparent magnitudes. Our proposed method allowed us to find such a correlation. This demonstrates that further research in this area is needed.

\scriptsize
\bibliographystyle{ComBAO}
\nocite{*}
\bibliography{references}

@INPROCEEDINGS{Ambartsumian1958,
       author = {{Ambartsumian}, V.~A.},
        title = "{On the Evolution of Galaxies}",
    booktitle = {La structure et l'{\'e}volution de l'universe},
         year = 1958,
       editor = {{Oort}, J.},
        month = jan,
        pages = {241-249},
       adsurl = {https://ui.adsabs.harvard.edu/abs/1958seu..conf..241A}
}

@ARTICLE{Ambartsumian1961,
       author = {{Ambartsumian}, V.~A.},
        title = "{Instability phenomena in systems of galaxies.}",
      journal = {\aj},
         year = 1961,
        month = dec,
       volume = {66},
        pages = {536-540},
          doi = {10.1086/108460},
       adsurl = {https://ui.adsabs.harvard.edu/abs/1961AJ.....66..536A}
}

@ARTICLE{Binggeli1993,
       author = {{Binggeli}, B. and {Popescu}, C.~C. and {Tammann}, G.~A.},
        title = "{The kinematics of the Virgo cluster revisited.}",
      journal = {Astron. Astroph. Suppl.},
         year = 1993,
        month = apr,
       volume = {98},
        pages = {275},
       adsurl = {https://ui.adsabs.harvard.edu/abs/1993A&AS...98..275B}
}

@ARTICLE{Binggeli1987,
       author = {{Binggeli}, Bruno and {Tammann}, G.~A. and {Sandage}, Allan},
        title = "{Studies of the Virgo Cluster. VI. Morphological and Kinematical Structure of the Virgo Cluster}",
      journal = {Astron. J},
         year = 1987,
        month = aug,
       volume = {94},
        pages = {251},
          doi = {10.1086/114467},
       adsurl = {https://ui.adsabs.harvard.edu/abs/1987AJ.....94..251B}
}

@ARTICLE{Binggeli1985,
       author = {{Binggeli}, B. and {Sandage}, A. and {Tammann}, G.~A.},
        title = "{Studies of the Virgo cluster. II. A catalog of 2096 galaxies in the Virgo cluster area.}",
      journal = {Astron. J},
         year = 1985,
        month = sep,
       volume = {90},
        pages = {1681-1758},
          doi = {10.1086/113874},
       adsurl = {https://ui.adsabs.harvard.edu/abs/1985AJ.....90.1681B}
}

@ARTICLE{Bohringer1994,
       author = {{B{\"o}hringer}, H. and {Briel}, U.~G. and {Schwarz}, R.~A. and {Voges}, W. and {Hartner}, G. and {Tr{\"u}mper}, J.},
        title = "{The structure of the Virgo cluster of galaxies from Rosat X-ray images}",
      journal = {Nature},
         year = 1994,
        month = apr,
       volume = {368},
       number = {6474},
        pages = {828-831},
          doi = {10.1038/368828a0},
       adsurl = {https://ui.adsabs.harvard.edu/abs/1994Natur.368..828B}
}

@ARTICLE{Dickey1994,
       author = {{Dickey}, J.~O. and {Bender}, P.~L. and {Faller}, J.~E. and {Newhall}, X.~X. and {Ricklefs}, R.~L. and {Ries}, J.~G. and {Shelus}, P.~J. and {Veillet}, C. and {Whipple}, A.~L. and {Wiant}, J.~R. and {Williams}, J.~G. and {Yoder}, C.~F.},
        title = "{Lunar Laser Ranging: A Continuing Legacy of the Apollo Program}",
      journal = {Science},
         year = 1994,
        month = jul,
       volume = {265},
       number = {5171},
        pages = {482-490},
          doi = {10.1126/science.265.5171.482},
       adsurl = {https://ui.adsabs.harvard.edu/abs/1994Sci...265..482D}
}

@ARTICLE{Dressler1980,
       author = {{Dressler}, A.},
        title = "{Galaxy morphology in rich clusters: implications for the formation and evolution of galaxies.}",
      journal = {Astroph. J},
         year = 1980,
        month = mar,
       volume = {236},
        pages = {351-365},
          doi = {10.1086/157753},
       adsurl = {https://ui.adsabs.harvard.edu/abs/1980ApJ...236..351D}
}

@ARTICLE{Fadda1998,
       author = {{Fadda}, D. and {Slezak}, E. and {Bijaoui}, A.},
        title = "{Density estimation with non-parametric methods}",
      journal = {Astron. Astroph. Suppl.},
         year = 1998,
        month = jan,
       volume = {127},
        pages = {335-352},
          doi = {10.1051/aas:1998355},
archivePrefix = {arXiv},
       eprint = {astro-ph/9704096},
 primaryClass = {astro-ph},
       adsurl = {https://ui.adsabs.harvard.edu/abs/1998A&AS..127..335F}
}

@ARTICLE{Hartog1996,
       author = {{den Hartog}, R. and {Katgert}, P.},
        title = "{On the dynamics of the cores of galaxy clusters.}",
      journal = {Mont. Not. R. Astron. Soc.},
         year = 1996,
        month = mar,
       volume = {279},
       number = {2},
        pages = {349-388},
          doi = {10.1093/mnras/279.2.349},
       adsurl = {https://ui.adsabs.harvard.edu/abs/1996MNRAS.279..349D}
}

@ARTICLE{Harutyunian2008,
       author = {{Harutyunian}, H.~A.},
        title = "{On the nature of cD galaxies}",
      journal = {Astrophysics},
         year = 2008,
        month = apr,
       volume = {51},
       number = {2},
        pages = {141-152},
          doi = {10.1007/s10511-008-9014-8},
       adsurl = {https://ui.adsabs.harvard.edu/abs/2008Ap.....51..141H}
}

@ARTICLE{Harutyunian2021,
       author = {{Harutyunian}, H.~A.},
        title = "{Discrepancy between Values of the Hubble Constant Obtained by Different Methods}",
      journal = {Astrophysics},
         year = 2021,
        month = dec,
       volume = {64},
       number = {4},
        pages = {435-445},
          doi = {10.1007/s10511-021-09705-z},
       adsurl = {https://ui.adsabs.harvard.edu/abs/2021Ap.....64..435H}
}

@ARTICLE{Harutyunian2024,
       author = {{Harutyunian}, H.~A.},
        title = "{Dark matter in the presence of dark energy}",
      journal = {Astrophysics},
         year = 2024,
        month = dec,
       volume = {67},
       number = {4},
        pages = {506-519},
          doi = {10.1007/s10511-025-09848-3}, 
       adsurl = {https://ui.adsabs.harvard.edu/abs/2024Ap.....67..506H}
}

@ARTICLE{Harutyunian2019,
       author = {{Harutyunian}, H.~A. and {Grigoryan}, A.~M. and {Khasawneh}, A.},
        title = "{On the correlation between average velocities of galaxies and their average luminosities in the closest large clusters of galaxies}",
      journal = {Communications of the Byurakan Astrophysical Observatory},
         year = 2019,
        month = sep,
       volume = {66},
       number = {1},
        pages = {25-30},
          doi = {10.52526/25792776-2019.66.1-25},
       adsurl = {https://ui.adsabs.harvard.edu/abs/2019CoBAO..66...25H}
}

@ARTICLE{Krasinsky2004,
       author = {{Krasinsky}, G.~A. and {Brumberg}, V.~A.},
        title = "{Secular increase of astronomical unit from analysis of the major planet motions, and its interpretation}",
      journal = {Celestial Mechanics and Dynamical Astronomy},
         year = 2004,
        month = nov,
       volume = {90},
       number = {3-4},
        pages = {267-288},
          doi = {10.1007/s10569-004-0633-z},
       adsurl = {https://ui.adsabs.harvard.edu/abs/2004CeMDA..90..267K}
}

@ARTICLE{Nolthenius1993,
       author = {{Nolthenius}, Richard},
        title = "{A Revised Catalog of CfA1 Galaxy Groups in the Virgo/Great Attractor Flow Field}",
      journal = {Astrophys.J.Suppl.},
         year = 1993,
        month = mar,
       volume = {85},
        pages = {1},
          doi = {10.1086/191753},
       adsurl = {https://ui.adsabs.harvard.edu/abs/1993ApJS...85....1N}
}

@ARTICLE{Perlmutter1999,
       author = {{Perlmutter}, S. and {Aldering}, G. and {Goldhaber}, G. and {Knop}, R.~A. and {Nugent}, P. and {Castro}, P.~G. and {Deustua}, S. and {Fabbro}, S. and {Goobar}, A. and {Groom}, D.~E. and {Hook}, I.~M. and {Kim}, A.~G. and {Kim}, M.~Y. and {Lee}, J.~C. and {Nunes}, N.~J. and {Pain}, R. and {Pennypacker}, C.~R. and {Quimby}, R. and {Lidman}, C. and {Ellis}, R.~S. and {Irwin}, M. and {McMahon}, R.~G. and {Ruiz-Lapuente}, P. and {Walton}, N. and {Schaefer}, B. and {Boyle}, B.~J. and {Filippenko}, A.~V. and {Matheson}, T. and {Fruchter}, A.~S. and {Panagia}, N. and {Newberg}, H.~J.~M. and {Couch}, W.~J. and {Project}, The Supernova Cosmology},
        title = "{Measurements of {\ensuremath{\Omega}} and {\ensuremath{\Lambda}} from 42 High-Redshift Supernovae}",
      journal = {Astroph. J},
         year = 1999,
        month = jun,
       volume = {517},
       number = {2},
        pages = {565-586},
          doi = {10.1086/307221},
archivePrefix = {arXiv},
       eprint = {astro-ph/9812133},
 primaryClass = {astro-ph},
       adsurl = {https://ui.adsabs.harvard.edu/abs/1999ApJ...517..565P}
}

@ARTICLE{Pinkney1993,
       author = {{Pinkney}, J. and {Rhee}, G. and {Burns}, J.~O. and {Hill}, J.~M. and {Oegerle}, W. and {Batuski}, D. and {Hintzen}, P.},
        title = "{The Dynamics of the Galaxy Cluster Abell 2634}",
      journal = {Astroph. J},
         year = 1993,
        month = oct,
       volume = {416},
        pages = {36},
          doi = {10.1086/173213},
       adsurl = {https://ui.adsabs.harvard.edu/abs/1993ApJ...416...36P}
}

@ARTICLE{Riess1998,
       author = {{Riess}, Adam G. and {Filippenko}, Alexei V. and {Challis}, Peter and {Clocchiatti}, Alejandro and {Diercks}, Alan and {Garnavich}, Peter M. and {Gilliland}, Ron L. and {Hogan}, Craig J. and {Jha}, Saurabh and {Kirshner}, Robert P. and {Leibundgut}, B. and {Phillips}, M.~M. and {Reiss}, David and {Schmidt}, Brian P. and {Schommer}, Robert A. and {Smith}, R. Chris and {Spyromilio}, J. and {Stubbs}, Christopher and {Suntzeff}, Nicholas B. and {Tonry}, John},
        title = "{Observational Evidence from Supernovae for an Accelerating Universe and a Cosmological Constant}",
      journal = {Astron. J},
         year = 1998,
        month = sep,
       volume = {116},
       number = {3},
        pages = {1009-1038},
          doi = {10.1086/300499},
archivePrefix = {arXiv},
       eprint = {astro-ph/9805201},
 primaryClass = {astro-ph},
       adsurl = {https://ui.adsabs.harvard.edu/abs/1998AJ....116.1009R}
}

@ARTICLE{Salvador-Sole1993,
       author = {{Salvador-Sole}, Eduard and {Gonzalez-Casado}, Guillermo and {Solanes}, Jose M.},
        title = "{Small-Scale Substructure in Relaxed Clusters. I. Statistical Characterization}",
      journal = {Astroph. J},
         year = 1993,
        month = jun,
       volume = {410},
        pages = {1},
          doi = {10.1086/172718},
       adsurl = {https://ui.adsabs.harvard.edu/abs/1993ApJ...410....1S}
}

@ARTICLE{Serna1996,
       author = {{Serna}, A. and {Gerbal}, D.},
        title = "{Dynamical search for substructures in galaxy clusters. A hierarchical clustering method.}",
      journal = {Astron. Astrophys.},
         year = 1996,
        month = may,
       volume = {309},
        pages = {65-74},
          doi = {10.48550/arXiv.astro-ph/9509080},
archivePrefix = {arXiv},
       eprint = {astro-ph/9509080},
 primaryClass = {astro-ph},
       adsurl = {https://ui.adsabs.harvard.edu/abs/1996A&A...309...65S}
}

@ARTICLE{Solanes1999,
       author = {{Solanes}, Jos{\'e} M. and {Salvador-Sol{\'e}}, Eduard and {Gonz{\'a}lez-Casado}, Guillermo},
        title = "{Substructure in the ENACS clusters}",
      journal = {Astron. Astrophys.},
         year = 1999,
        month = mar,
       volume = {343},
        pages = {733-739},
          doi = {10.48550/arXiv.astro-ph/9812103},
archivePrefix = {arXiv},
       eprint = {astro-ph/9812103},
 primaryClass = {astro-ph},
       adsurl = {https://ui.adsabs.harvard.edu/abs/1999A&A...343..733S}
}

@ARTICLE{Vaucouleurs1979,
       author = {{de Vaucouleurs}, G. and {Nieto}, J.-L.},
        title = "{A photometric analysis of the jet in Messier 87.}",
      journal = {Astrophys. J},
         year = 1979,
        month = jul,
       volume = {231},
        pages = {364-371},
          doi = {10.1086/157199},
       adsurl = {https://ui.adsabs.harvard.edu/abs/1979ApJ...231..364D}
}

@ARTICLE{Harutyunian2022,
       author = {{Harutyunian}, H.~A.},
        title = "{Can the Existence of Dark Energy Shed Light on the Dark Sides of the ``Byurakan Concept''?}",
      journal = {Communications of the Byurakan Astrophysical Observatory},
         year = 2022,
        month = aug,
       volume = {69},
       number = {1},
        pages = {1-9},
          doi = {10.52526/25792776-22.69.1-1},
       adsurl = {https://ui.adsabs.harvard.edu/abs/2022CoBAO..69....1H}
}

@ARTICLE{Zwicky1933,
       author = {{Zwicky}, F.},
        title = "{Die Rotverschiebung von extragalaktischen Nebeln}",
      journal = {Helvetica Physica Acta},
         year = 1933,
        month = jan,
       volume = {6},
        pages = {110-127},
       adsurl = {https://ui.adsabs.harvard.edu/abs/1933AcHPh...6..110Z}
}

@ARTICLE{Zwicky1937ApJ,
       author = {{Zwicky}, F.},
        title = "{On the Masses of Nebulae and of Clusters of Nebulae}",
      journal = {\apj},
         year = 1937,
        month = oct,
       volume = {86},
        pages = {217},
          doi = {10.1086/143864},
       adsurl = {https://ui.adsabs.harvard.edu/abs/1937ApJ....86..217Z}
}

@ARTICLE{Lainey2020a,
       author = {{Lainey}, Val{\'e}ry and {Casajus}, Luis Gomez and {Fuller}, Jim and {Zannoni}, Marco and {Tortora}, Paolo and {Cooper}, Nicholas and {Murray}, Carl and {Modenini}, Dario and {Park}, Ryan S. and {Robert}, Vincent and {Zhang}, Qingfeng},
        title = "{Resonance locking in giant planets indicated by the rapid orbital expansion of Titan}",
      journal = {Nature Astronomy},
         year = 2020,
        month = jun,
       volume = {4},
        pages = {1053-1058},
          doi = {10.1038/s41550-020-1120-5},
archivePrefix = {arXiv},
       eprint = {2006.06854},
 primaryClass = {astro-ph.EP},
       adsurl = {https://ui.adsabs.harvard.edu/abs/2020NatAs...4.1053L}
}

@ARTICLE{Lainey2020b,
       author = {{Lainey}, Val{\'e}ry and {Casajus}, Luis Gomez and {Fuller}, Jim and {Zannoni}, Marco and {Tortora}, Paolo and {Cooper}, Nicholas and {Murray}, Carl and {Modenini}, Dario and {Park}, Ryan S. and {Robert}, Vincent and {Zhang}, Qingfeng},
        title = "{Publisher Correction: Resonance locking in giant planets indicated by the rapid orbital expansion of Titan}",
      journal = {Nature Astronomy},
         year = 2020,
        month = jun,
       volume = {4},
        pages = {809-809},
          doi = {10.1038/s41550-020-1149-5},
       adsurl = {https://ui.adsabs.harvard.edu/abs/2020NatAs...4..809L}
}

\end{document}